\documentclass[twocolumn,aps,prd,nofootinbib]{revtex4-1}
\usepackage[T1]{fontenc}
\usepackage[latin1]{inputenc}
\usepackage{amssymb}
\usepackage{amsmath}
\usepackage{enumerate}
\usepackage{graphicx,wasysym}
\usepackage[pdftex]{hyperref}

\def\K{K{\"a}hler}

\makeatletter

\usepackage{epsf}

\def\be{\begin{equation}}
\def\ee{\end{equation}}
\def\ba{\begin{eqnarray}}
\def\ea{\end{eqnarray}} 
\def\beas{\begin{eqnarray*}}
\def\eeas{\end{eqnarray*}}
\def\sla{\raise.15ex\hbox{$/$}\kern-.57em}

\begin{document}

\rightline{LPT-Orsay 11/89, UMN--TH--3018/11, FTPI--MINN--11/26}

\title{\Large\bf   Supersymmetry Breaking due to Moduli Stabilization in String Theory}

\author{Andrei Linde$^{1}$}
\author{Yann Mambrini$^{2}$}
\author{Keith A. Olive$^{3}$}

\affiliation{$^{1}$Stanford Institute of Theoretical Physics and Department of Physics, Stanford University, Stanford, CA 94305, USA\\
$^{2}$Laboratoire de Physique Th\'eorique, Universit\'e Paris-Sud, F91405 Orsay, France \\
$^{3}$William I. Fine Theoretical Physics Institute, School of Physics and Astronomy, University of Minnesota, Minneapolis, MN 55455, USA}

\begin{abstract}

We consider the phenomenological consequences of fixing compactification moduli.
In the simplest KKLT constructions, stabilization of internal dimensions is rather soft:  
weak scale masses for moduli are generated, and are of order $m_\sigma \sim m_{3/2}$.  As
a consequence one obtains a pattern of soft supersymmetry breaking masses 
found in gravity and/or anomaly mediated supersymmetry breaking (AMSB) models.  
These models may lead to destabilization of internal dimensions in the early universe, 
unless the Hubble constant during inflation is very small. 
Fortunately, strong stabilization of compactified dimensions can be achieved by a proper choice of the superpotential 
(e.g in the KL model with a racetrack superpotential). This allows for a solution of the cosmological moduli problem and for a successful implementation of inflation in supergravity.  We show that strong moduli stabilization leads to a very distinct pattern of soft supersymmetry breaking masses. In general, we find that soft scalar masses remain of order the gravitino mass, while gaugino masses nearly vanish at the tree level, i.e. they are of order $m_{3/2}^2/m_\sigma$. Radiative corrections generate contributions to gaugino masses reminiscent of AMSB models and a decoupled spectrum of scalars reminiscent of split-supersymmetry. This requires a relatively large gravitino mass ($\sim \mathcal{O}(100)$ TeV), resolving the cosmological gravitino problem and problems with tachyonic staus in AMSB models.

\end{abstract}

\maketitle

\section{Introduction}

Soon after the introduction of $N=1$ supergravity as a phenomenological model \cite{pol,sugra,nilles}
it was recognized that there is a cosmological moduli problem~\cite{Polonyi}. The problem is manifest 
whenever there is a weakly (gravitationally) coupled field with a weak scale $\sim \mathcal{O}(1)$ TeV mass and a Planck scale vacuum expectation value.  This field may either be related to the mechanism of supersymmetry breaking as in the case of the Polonyi potential~\cite{pol}, 
or associated with 
one of many scalars in string theory such as compactification moduli.

In any attempt to resolve the moduli problem, one must first tackle the problem of moduli stabilization.
An attractive and often studied mechanism for the vacuum stabilization of moduli is the KKLT mechanism~\cite{Kachru:2003aw}. The KKLT mechanism consists of two basic steps. First of all, one finds a stable supersymmetric
anti-deSitter (AdS) vacuum, then this vacuum is uplifted to a de Sitter vacuum with an extremely small positive cosmological constant.
One may interpret uplifting either as soft supersymmetry breaking induced by string theory effects \cite{Kachru:2003aw}, or as a D-term contribution 
\cite{Burgess:2003ic,Achucarro:2006zf,Dudas:2006vc}, or an F-term contribution in N = 1 supergravity   \cite{Kallosh:2006dv}.
The phenomenology of this class of models has been heavily studied (see e.g. \cite{Choi:2004sx,
Choi:2005ge,Endo:2005uy,Choi:2005uz,Falkowski:2005ck,Lebedev:2006qc}). Soft supersymmetry breaking masses arise through a combination
 of modulus (gravity) and anomaly mediation leading to a pattern of masses with relatively heavy gravitino ($\sim$ 100 TeV) and a compressed
 scalar-gaugino spectrum around the TeV scale.

However, unless the moduli are very heavy, the cosmological problems associated with their
late-time evolution persist. In addition, these models may suffer from a vacuum destabilization 
during inflation unless the Hubble parameter at the last stage of inflation is smaller than the gravitino mass ~\cite{Kallosh:2004yh}.
In a generalization of the KKLT model based on a racetrack superpotential, also known 
as the KL model \cite{Kallosh:2004yh}, the problem of destabilization is resolved as the moduli in this model are
superheavy and allow for significantly higher scales for the Hubble parameter and a
successful implementation of the chaotic inflationary scenario in supergravity \cite{klor}.
The KL model, however, leads to a very distinct pattern of soft supersymmetry breaking masses
with gaugino masses typically much lighter than scalar masses as in models of split supersymmetry breaking \cite{split}.  
We show that this result is a general consequence of any model in which the 
moduli are much heavier than the gravitino.

In what follows, we will first describe in Section II the basic ingredients of the KKLT and KL models.
We will restrict our attention here to supersymmetry breaking via uplifting.
We will also restrict our attention to a minimal \K\ potential for Standard Model (SM) fields,
though we will comment on the consequences of generalizing to non-minimal models. 
In section III,  we compare the contributions to soft masses in both the KKLT and KL models. 
We will show that  in the simplest KKLT models,  gaugino masses are proportional to the gravitino mass, while
in KL models, they are proportional to $m_{3/2}^2/m_{\sigma}$. Here $m_{\sigma}$ is the mass of the volume modulus, describing volume stabilization in string theory; $m_{\sigma} \gg m_{3/2}$ in the KL scenario. In both classes of models, 
soft scalar masses are set by $m_{3/2}$, but in the KL model the gaugino masses practically vanishes unless one takes into account quantum corrections. As a consequence, anomaly mediation becomes the first
source of breaking terms in the gaugino sector of the theory.
In an appendix, we show that this result is not restricted to the KL model, 
but is a general consequence of any model with a strongly stabilized vacuum, 
such that  $m_\sigma \gg m_{3/2}$.
Phenomenological implications of these results and our conclusions
are summarized in section IV.

\section{The KKLT and KL models}\label{generalpot}
In this section, we will briefly review the KKLT and KL models. We will couple the KKLT(KL) sector
to the SM through gravity and assume minimal $N=1$ supergravity for the SM sector as a starting point.  
The KKLT (KL) sector consists of a single chiral field: the modulus $\rho$, and we will denote SM fields collectively as $y$. 
 The scalar potential for uncharged chiral superfields in $\mathcal N=1$ supergravity is
\ba
V &=& e^{ K}\left({ K}^{i\bar j}D_i W\bar D_{\bar j}\bar{ W}-3| W|^2\right)
\nonumber
\\
&=&e^G \left( G_i G^{i \bar j}G_{\bar j} -3 \right), 
\label{eqn:SUGRApotential}
\ea
with
\be
D_i W\equiv\partial_i W+{ K}_iW ~~~\mathrm{and}~~~ G = { K} + \log |{ W}|^2 .
\ee

In string theory, one must consider stabilization of the volume modulus $\rho$ to explain why our universe is 4d rather than 10d. The simplest approach to this issue is based on the KKLT mechanism \cite{Kachru:2003aw}. 
As we will see, in this theory, one first finds a stable supersymmetric vacuum with a negative vacuum energy density $V_{\rm AdS}$, and then uplifts it until its vacuum energy becomes positive but negligibly small, about $10^{{-120}}$ in Planck units. After uplifting, supersymmetry is broken, and the gravitino mass has a simple relation to the depth of the original AdS minimum  \cite{Kallosh:2004yh}:
\be
m^{2}_{3/2} = |V_{\rm AdS}|/3 \ .
\label{m32ads}
\ee
Thus the mechanism of supersymmetry breaking is built into the new generation of string theory models. One can add to it other mechanisms of supersymmetry breaking, such the Polonyi mechanism \cite{pol}, dynamical supersymmetry breaking \cite{Intriligator:2006dd}, an O'Raifeartaigh mechanism \cite{O'Raifeartaigh:1975pr,Kallosh:2006dv} or something else. However, this would make the models more complicated.  Therefore, in this paper we will concentrate on the string theory based mechanism of supersymmetry breaking.

For both the KKLT and KL models, we 
define a \K\ potential as
\be
K = -3 \log (\rho + \bar{\rho}) + y \bar{y} \ .
\label{kahl}
\ee
In each case we will assume the superpotential is separable so that
\be
W = W(\rho) + W_{SM}(y)
\label{super}
\ee
where $W(\rho)$ is either the KKLT or KL superpotential and 
$W_{SM}(y)$ is the superpotential for the Standard Model (the subscript SM can
be dropped without loss of clarity). 
From Eqs. (\ref{eqn:SUGRApotential}) and (\ref{super}),
we can write the scalar potential as
\be
V = e^K (|D_\rho W|^2 + |D_y W|^2 - 3 |W(\rho)+ W(y)|^2 )\ .
\label{fullpot}
\ee
Here $|D_\rho W|^2 =K^{\rho \bar{\rho}} D_\rho W {\bar D}_{\bar\rho} \bar W$. In the remainder of this section, we will ignore the SM contributions to the
scalar potential.
Along the direction $\sigma = {\rm Re}\, \rho$, we can write
\be
\partial_{\sigma}V =
 \frac{D_{\rho}{W}}{3\sigma^2} (\sigma W_{\rho,\rho} - 2W_{\rho}) \ ,
 \label{partialV}
\ee
this equation being valid for any real superpotential $W$.

The KKLT model is specified by the choice of superpotential
\be
W_{KKLT} = W_0 + Ae^{-a\rho} \ ,
\label{kklt}
\ee
where $W_0$ and $a>0$ are constants. 
In this theory, there is a supersymmetry preserving AdS minimum
found by setting $D_\rho W = 0$. Since $K_\rho = -3/(2\sigma)$,
we can use $D_\rho W = 0$ to solve for $W_0$ in terms of  $\rho = \sigma_0$ corresponding to the 
minimum of the potential,
\be
W_0 = -A e^{-a \sigma_0} \left(1+\frac{2}{3} a \sigma_0 \right) .
\ee
Furthermore, at the minimum, 
\be\label{W0}
W(\sigma_0) =- \frac{2 a \sigma_0}{3} A e^{-a\sigma_0} \qquad W_\rho(\sigma_0) = - a A e^{-a\sigma_0} ,
\ee
and the scalar potential at the minimum becomes
\be\label{ADSV}
V_{\rm AdS} = V(\sigma_0) = - \frac{a^2 A^2 e^{-2a \sigma_0}}{6\sigma_0} .
\ee
Using the relation $m_{3/2}^2 = e^G = e^K \left| W \right|^2$, one finds
\be\label{gravmass}
m_{3/2}=\frac{aA}{3(2 \sigma_0)^{1/2}}e^{-a \sigma_0} \ ,
\ee
in agreement with Eq. (\ref{m32ads}). As we will see, this relation remains valid after the uplifting of the AdS minimum.

Note that the validity of the KKLT model requires that $\sigma_{0}\gg 1$ and $a\sigma_{0} > 1$  \cite{Kachru:2003aw}.   In the following, we will assume that  $a\sigma_{0} \gg 1$, to simplify the analytical calculations. In fact, this condition must be satisfied if we want to find the gravitino mass in the TeV range, which implies that $m_{3/2} \sim 10^{{-15}}$ in Planck mass units. Indeed, the parameters $a$ and $A$ of the KKLT construction are not expected to be exponentially small. Therefore, according to (\ref{gravmass}), the only way to have a hierarchically small gravitino mass $m_{3/2} \sim 10^{{-15}} \sim e^{-35}$ is to have $a\sigma_{0} \gg 1$. For example, for $a \sim A \sim 1$ and $m_{3/2} \sim 10^{-15}$,  Eq. (\ref{gravmass}) yields $\sigma_{0} \sim  a \sigma_{0} \approx 30$.

It is useful to represent the expressions for $m_{3/2}$ in a different but equivalent way. Using (\ref{W0}), one finds, for $a\sigma_{0}\gg 1$,
\be\label{gravmass2}
m_{3/2}=\frac{W_{0}}{(2\sigma_0)^{3/2}}  \ .
\ee
In the example considered above one finds $W_{0}  \sim 10^{-12}$. This is the same degree of fine-tuning as the one required in the standard Polonyi model.

To find the relation between the mass of the volume modulus and the gravitino mass, 
one can can use Eq. (\ref{partialV}), which yields
\be
m_{\sigma} ^2 = \frac{\partial_{\sigma}\partial_{\sigma}V}{2K_{\rho,\bar\rho}} =
 \frac{2}{9}(D_{\rho}{W})_\sigma(\sigma W_{\rho,\rho} - 2W_{\rho}) 
 \label{partialpartialV}
\ee
at the supersymmetric minimum.
 In this case, one can
show that the mass of the volume modulus, as well as the mass of its imaginary (axionic) component,  is given by
\be\label{modmass}
m_{\sigma}  = \frac{\sqrt{2 \sigma_{0}}}{3} W_{\rho,\rho} =  2 a\sigma_{0}\, m_{3/2} \ .
\ee
 As a result, the mass of the volume modulus is not much greater than the gravitino mass. And this means that the volume stabilization in the KKLT scenario is actually very soft \cite{Kallosh:2004yh}.

\begin{figure}
\centering
\includegraphics[scale=0.82]{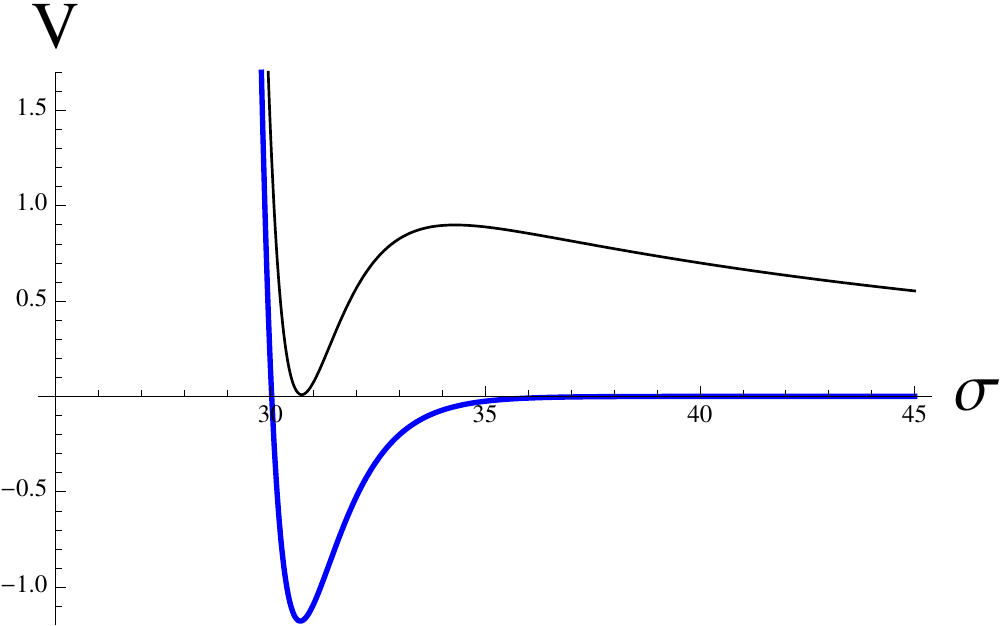}
\caption{Scalar potential of the KKLT model for the values of the parameters $A =1$, $a = 1$ and $W_0=10^{-12}$  before and after uplifting.
The potential has been multiplied by a factor of $10^{29}$ for clarity.}
\label{KKLTpotential}
\end{figure}

As mentioned above, the AdS minimum at $\rho = \sigma_0$ must be uplifted. We illustrate the uplifting effect in the Fig.~\ref{KKLTpotential} in the phenomenologically reasonable case
$A=1$, $a=1$ and $W_0=10^{-12}$ corresponding to a gravitino mass $m_{3/2} = 2 \times 10^{-15}$.
The position of the minimum in this case is $\sigma_0 \approx 30.7$. 

Uplifting of the AdS minimum induces supersymmetry breaking and is achieved by adding to the potential a term 
\be
\Delta V \approx |V_{\rm AdS}|\ {\sigma^{n}_{0}\over\sigma^{n}} \ .
\label{deltaV}
\ee
 In the original KKLT construction it was assumed that $n = 3$ \cite{Kachru:2003aw}, but according to \cite{Kachru:2003sx} $n = 2$ in
  the uplifting term, due to effects related to warping. One may have $n = 3$ if the uplifting occurs due to a 
  D-term \cite{Burgess:2003ic,Achucarro:2006zf,Dudas:2006vc}. The choice of $n$ will not affect our qualitative conclusions; for definiteness, we will choose $n=2$. 
  
Because of the dependence of the uplifting term on $\sigma$, the minimum after the uplifting shifts to greater values of $\sigma$. Let us denote the shift in $\sigma$ after uplifting by $\Delta \sigma$. 
Once again, using Eq. (\ref{partialV}), we can obtain $\Delta \sigma$ from
\be
(D_{\rho} W)_\sigma (\sigma_0 W_{\rho,\rho} - 2W_{\rho}) \Delta \sigma = 6 V_{\rm AdS} \sigma_0 \ .
\label{solvekklt2}
\ee
Then, for $a\sigma_{0} \gg 1$, one can show that the relative value of the shift is small,
\be
{\Delta \sigma\over \sigma_{0}} = (a\sigma_{0})^{{-2}} \ll 1.
\ee
This shift does not change much the values of the superpotential and the \K\ potential, which depend on $\sigma$. Therefore the gravitino mass after the uplifting is correctly represented by the expression (\ref{gravmass}). Similarly, the mass of the volume modulus (\ref{modmass}) practically does not change during the uplifting.

In  $N = 1$ supergravity, ignoring D-term and string theory effects leading to the uplifting, one could use the standard expression for the potential
$V = e^{K}(|D_{\rho}W|^{2} - 3|W|^{2})$.
Then the existence of the gravitino mass $m^{2}_{3/2} = e^{K} |W|^{2}$ in a Minkowski vacuum with $V = 0$ would automatically imply that $|D_{\rho}W|^{2} = 3e^{-K}  m^{2}_{3/2} \not = 0$.

 However, because of uplifting, this relation is no longer valid. Since $D_\rho W = 0$ in the supersymmetric AdS vacuum, the value of $D_\rho W$ after uplifting is completely determined by the small shift $\Delta \sigma$. For  $a\sigma_{0} \gg 1$, it is given by
\be\label{KKLTDW}
D_\rho W = (D_{\rho} W)_\sigma \Delta \sigma \simeq W_{\rho,\rho} \Delta \sigma =  {3\sqrt 2\over a \sqrt \sigma_{0}}\, m_{3/2} \ .
\ee
As we see, all of the important parameters, such as $W$, $W_\rho$, $D_\rho W$, $m_{\sigma}$, have their scale determined by the gravitino mass, up to some combination of parameters $a$ and $\sigma_{0}$. This fact will be important for us in the next section when we will discuss consequences of the KKLT mechanism for supersymmetry breaking in the observable sector.

The softness of the moduli stabilization in the simplest versions of the KKLT construction leads to a rather unusual problem:
 the Hubble constant during inflation cannot be much greater than the gravitino mass, $H \lesssim m_{3/2}$  \cite{Kallosh:2004yh}. The reason is that in the simplest KKLT models, the barrier separating the stabilized dS vacuum from the 10d Minkowski vacuum has a height proportional to $m^{2}_{3/2}$. When the inflationary potential is added to the system, it may lift the dS minimum above the barrier. If this happens, the universe decompactifies and becomes 10-dimensional.  Thus, for $m_{3/2} \lesssim $ 1 TeV, one must have a very low value of the Hubble constant at the last stage of inflation in the KKLT based inflationary models. Special effort is required to build inflationary models of this type.

One can try to solve this problem in several different ways, see for example \cite{Kallosh:2004yh,Davis:2008fv,Badziak:2008yg,Conlon:2008cj,McAllister:2008hb}. The simplest mechanism involves a slightly generalized KKLT model, which is sometimes called the KL model \cite{Kallosh:2004yh}. In this model, 
instead of the standard KKLT superpotential (\ref{kklt}), one uses the racetrack superpotential 
\be
W_\text{KL} = W_0 + Ae^{-a\rho}- Be^{-b\rho} \ .
\label{adssup}
\ee
In contrast to the KKLT case, the new degree of freedom offered by $Be^{-b\rho}$ allows the new model to have a supersymmetric 
Minkowski solution. Indeed, 
for the particular choice of $W_0$,
\be\label{w0}
W_0= -A \left({a\,A\over
b\,B}\right)^{a\over b-a} +B \left ({a\,A\over b\,B}\right) ^{b\over b-a} ,
\ee
the potential of the field $\sigma$ has a supersymmetric minimum with
$W(\sigma_{0})=0$,  $D_\rho W(\sigma_{0}) = 0$,   and $V(\sigma_{0})=0$,
where 
$
 \sigma_{0}= {1\over a-b}\ln \left ({a\,A\over b\,B}\right)\, .
$
In further contrast with the KKLT model, in the KL model there exist another minimum deeper than the supersymmetric one, implying that the SUSY minimum is metastable.

The gravitino mass in the supersymmetric Minkowski minimum vanishes, whereas  the mass squared of the field $\sigma$ at the minimum as well as the mass squared of the imaginary component of the field $\rho$,  is given by  \cite{klor}
\be
m^{2}_{\sigma} = \frac{2}{9} W_{\rho,\rho}^2 \sigma_0 = \frac{2}{9} a\, A\, b\, B\, (a-b) \left(\frac{a A}{b B}\right)^{-\frac{a+b}{a-b}}\,  \ln\left (\frac{a A}{b B}\right) \ .
\ee 
To understand the implications of this result, let us consider a particular simple choice of parameters 
$A=B =1,\,a = 0.1,\,b = 0.05$.
For these parameters, one has $m_{\sigma} \sim 2\times  10^{-3}$, in Planck units, so the field $\sigma$ is much heavier than the inflaton field, which, in the simplest model of chaotic inflation~\cite{Linde:1983gd} has mass $m_{\phi} \sim 6 \times 10^{{-6}}$. This hierarchy of mass scales is one of the necessary conditions which is required to ignore the dynamics of the volume modulus $\sigma$ during inflation.  What is most important here is that the mass $m_{\sigma}$, as well as the height of the barrier separating the Minkowski minimum from the AdS minimum in this model, does not have any relation to the gravitino mass. Therefore one can have inflation in this model for $H \gg m_{3/2}$ \cite{Kallosh:2004yh,Davis:2008fv,McAllister:2008hb,klor}.

In the KL model discussed so far, supersymmetry is unbroken in the vacuum state corresponding to the minimum of the potential with $V = 0$. The scale of  supersymmetry breaking  will be determined by a slight perturbation of the superpotential (\ref{adssup}) by adding to it a small constant $\Delta W$ proportional to the weak scale, $\mu$. Independent of the sign of $\Delta W$, the constant shifts the minimum of the potential $V$ from zero to a slightly negative value $V_{\rm AdS} < 0$ at $\sigma_0 + \delta \sigma$. Therefore $V_{\rm AdS}$ in the first approximation must be proportional to $-\Delta W^{2}$. At the shifted minimum, supersymmetry is preserved, and $D_\rho W (\sigma + \delta \sigma) = 0$.  
Since $W_\rho (\sigma_0)$ = 0, we can write $W_\rho (\sigma_0 + \delta \sigma) = W_{\rho,\rho}(\sigma_0) \delta \sigma$.
After some algebra, one finds that the position of the minimum shifts by 
$
\delta\sigma = {3\Delta W\over 2\sigma_{0} W_{\rho,\rho}},
$
and the potential at the minimum becomes \cite{klor}
\be
V_{\rm AdS}(\Delta W) = -{3(\Delta W)^2\over 8 \sigma_{0}^{3}}= - {3\over 8} \left({a - b \over  \ln \left({a A\over b B}\right)}\right)^{3}\, (\Delta W)^2 \ .
\label{vads}
\ee 
In this minimum, the value of the superpotential (including the additional constant $\Delta W$), is equal to $\Delta W$ up to small corrections $O(\Delta W)^2$. Supersymmetry in the minimum is still unbroken, $D_\rho W = 0$, whereas $W_{\rho} = {3\over 2 \sigma_0}\Delta W$. 
Note that the only ``large" quantities here (in Planck units) are $\sigma_0$ and $W_{\rho,\rho}$.
Therefore, $\delta \sigma \propto \Delta W \propto m_{3/2}$.

\begin{figure}[ht!]
\centering
\includegraphics[scale=0.82]{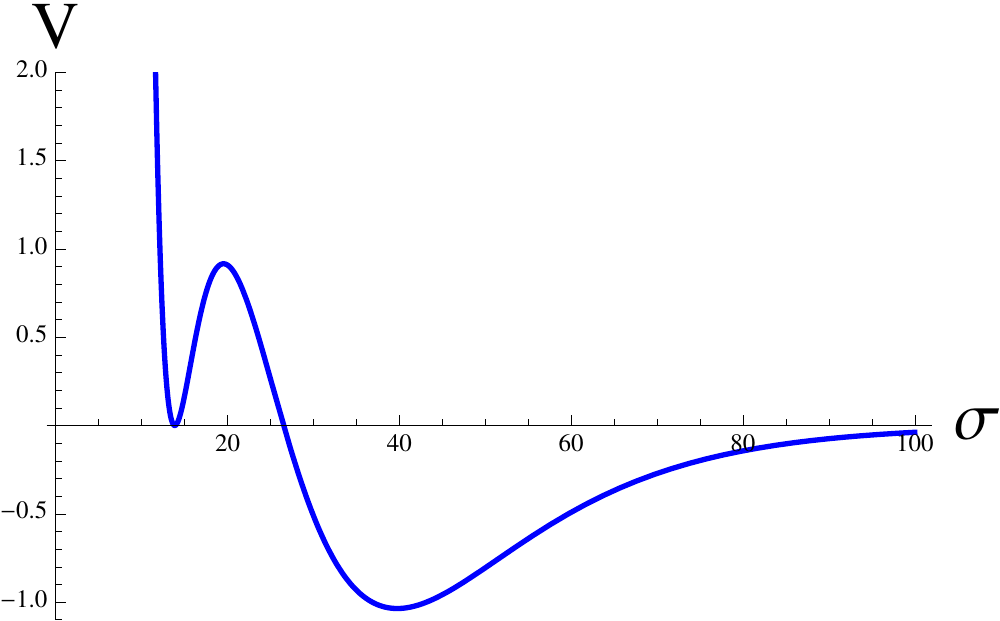}
\caption{Scalar potential of the KL model for the values of the parameters $A=B =1,\,a = 0.1,\,b = 0.05$.
The potential has been multiplied by a factor of $10^{7}$ for clarity. The effect of uplifting is so small as compared to the height of the barrier in this model that one cannot distinguish an uplifted and non-uplifted potential on the scale of this figure. 
}
\label{KLpotential}
\end{figure}

As in the case of the KKLT model,  we must add an uplifting potential.
After uplifting to the present state with a nearly vanishing vacuum energy, the gravitino mass  becomes 
\be
m_{3/2} = \sqrt{|V_{\rm AdS}|/3} =  {1\over 2\sqrt 2}\left({a - b \over  \ln \left({a A\over b B}\right)}\right)^{3/2}\, |\Delta W| \ .
\ee 
In particular, for $A=B =1,\,a = 0.1,\,b = 0.05$, one has $m_{3/2} \sim 7 \times 10^{-3} |\Delta W| \sim 3.5\, m_{{\sigma}} |\Delta W| \ll m_{{\sigma}}$.  
Since we assume $\Delta W$ to be very small (see below), we have $m_{3/2}/m_\sigma \ll 1$.
The shape of the potential, $V$, for this set of parameters is shown in Fig. \ref{KLpotential}. Note that the effect of uplifting is so tiny as compared to the height of the barrier in the KL model that one cannot distinguish an uplifted and non-uplifted potential on the scale of this figure. This helps to understand why the moduli stabilization in the KL model is so much stronger than in the simplest versions of the KKLT construction.

To have  $m_{{3/2}} \sim 1$~TeV $\sim 0.4\times 10^{{-15}}$ in Planck units in the model with $A=B =1,\,a = 0.1,\,b = 0.05$, one should have $|\Delta W| \sim 6\times 10^{-14}$. This means that to make the gravitino mass comparable to the electroweak scale, we must introduce a small parameter $\sim 10^{-13}$. This is the same degree of fine-tuning as the one required in the standard Polonyi model. Each of these models requires the presence of one very small number. 
In the KKLT model, it is $W_{0}$ that must be tuned small. In the KL model, we need
approximately the same small number added to a number of O(1) in the superpotential. In the context of our paper, it is important that the required accuracy of fine-tuning of the superpotential in the KL model is the same as in the simplest KKLT model, and in both cases the goal of fine-tuning is the same: to achieve a very low level of supersymmetry breaking.

Thus it is hard to discriminate between the simplest KKLT models with the superpotential (\ref{kklt}) and the KL models (\ref{adssup}) on the  basis of fine-tuning.
It would be interesting to check which one of these models is more probable in the landscape, but this would go beyond the scope of the present paper. The main advantage of the KL model is that the mass of the volume modulus can be many orders of magnitude greater than the gravitino mass. This strongly stabilizes the size of the compactified space and allows inflation with very high values of the Hubble constant, $H \gg m_{3/2}$ \cite{Kallosh:2004yh,Davis:2008fv,McAllister:2008hb,klor}. Moreover, in the KL model, the presence of the light Polonyi moduli fields is not required, which solves the cosmological moduli problem.  In addition, as we will see in the next section (see also \cite{klor}), this class of models provides a natural solution of the cosmological gravitino problem. 
Clearly, the KL model is not the only one where the strong moduli stabilization can be achieved. One may add
some new terms to the racetrack superpotential of the KL model, or one may find another, totally different superpotential which allows strong stabilization with $m_\sigma \gg m_{3/2}$. All of these models share certain features to be discussed below.

Uplifting in the KL model induces a shift in the modulus field,
which we will denote as $\Delta \sigma$ to distinguish it from the shift $\delta \sigma$
induced by adding the constant $\Delta W$ to the superpotential.
Eq. (\ref{solvekklt2}) remains valid in the KL model, and we can use it to find
$\Delta \sigma$. The left hand side of Eq. (\ref{solvekklt2}) should now be evaluated at 
$\sigma_0 + \delta \sigma$. Then since, $(D_\rho W)_\sigma  \approx W_{\rho,\rho} $, we can easily solve for the uplifting shift $\Delta \sigma = \frac{6 V_{\rm AdS}}{W_{\rho,\rho}^2} $.
One can now show that for any choice of the superpotential which leads to strong volume modulus stabilization with the \K\, potential (\ref{kahl}), the supersymmetry breaking term $D_\rho W$ in the minimum of the uplifted potential is given by
\be\label{KLDW}
D_\rho W = (D_{\rho} W)_\sigma \Delta \sigma \simeq W_{\rho,\rho} \Delta \sigma =   6 \sqrt{2\sigma_{0}}\,  \frac{m_{3/2}}{m_\sigma} \  m_{3/2}.
\ee
We give the derivation of this general result in the Appendix.  

The exact form of this equation will not be important for us. What is important is the following qualitative statement: In all models where the volume modulus can be strongly stabilized, i.e. $m_\sigma \gg m_{3/2}$, the supersymmetry breaking term $D_\rho W$ is strongly suppressed. The validity of this statement is very easy to understand: If the original AdS state is supersymmetric, the supersymmetry breaking term $D_\rho W$ vanishes before the uplifting. If the minimum is strongly stabilized (the potential sharply rises in the vicinity of the minimum), then adding a soft uplifting term   $\Delta V \approx |V_{\rm AdS}|\ {\sigma^{n}_{0}\over\sigma^{n}}$ barely affects the value of the field $\sigma$ in the minimum. As a result, uplifting in theories with strong volume modulus stabilization will keep the term $D_\rho W$ vanishingly small. As we will see, this conclusion may have important implications for particle phenomenology in the context of  models with strongly stabilized moduli.

\section{Soft masses in the KKLT and KL models}

We now return to Eqs. (\ref{kahl}) - (\ref{fullpot}) to consider the scalar potential of SM matter fields.
Inserting Eqs. (\ref{kahl}) and (\ref{super}) into (\ref{fullpot}),
we can write the scalar potential as
\begin{eqnarray}
&V  & =  \frac{e^{y^2}}{8 \sigma^3} \Bigl(-3  |W(\rho)+ W(y)|^2 
 +  |\bar{y} (W(\rho)+ W(y)) + W_y|^2  \nonumber \\
  &+ &
    \frac{4\sigma^2}{3}(D_\rho W (\rho) - \frac{3}{2\sigma} W(y))( {\bar D}_{\bar\rho} {\bar W} (\bar\rho) - \frac{3}{2\sigma} {\bar W}(\bar y)) \Bigr) ,
    \label{full}
\end{eqnarray}
assuming ${\rm Im}\, \rho = 0$. 
If we note first that we can remove contributions from  $\frac{4\sigma^2}{3} D_\rho W (\rho){\bar D}_{\bar\rho}  {\bar W} (\bar\rho) - 
3 |W(\rho)|^2$ as they will be cancelled by $V_{\rm uplift}$ and then expand the remaining
terms in
(\ref{full}) and take the low energy limit ($M_P \to \infty$),
we obtain,
\begin{eqnarray}
V  & = & \frac{1}{8 \sigma^3} \left(  |W_y|^2  + |W(\rho)|^2 y \bar{y}   \right. \nonumber \\
 & & + \left. \left( (y W_y - 3 W(y))\bar{W}(\rho) \right. \right. \nonumber \\
 && \left. \left. - 2 \sigma W(y) {\bar D}_{\bar\rho}  \bar{W}(\bar{\rho}) )+ h.c. \right)  \right)  .
    \label{lowfull}
\end{eqnarray}
If we rescale the superpotential by $W_{SM} \to 2\sqrt{2} \sigma^{3/2} W_{SM}$
(where $\sigma$ to be evaluated at the near-Minkowski minimum),
then we can write the potential in a more standard form  
\begin{eqnarray}
V_{SM} & = &   \left|\frac{\partial W_{SM}}{\partial y^i}\right|^2  +m_{0}^2 y^i \bar{y_i} + 
\left( A_0 W_{SM}+ h.c.\right)
\end{eqnarray}
where we have now defined the universal scalar mass as
\be
m_0^2 = \frac{1}{8\sigma^3} |W(\rho)|^2 \equiv m_{3/2}^2 ,
\label{scalar}
\ee
and a universal $A$-term
\be
A_0 W_{SM} = (y W_y - 3 W(y))m_{3/2} - \frac{1}{\sqrt{2\sigma}} {\bar D}_{\bar\rho}  \bar{W}(\bar{\rho}) W_{SM} .
\label{aterm}
\ee
Note that for tri-linears, the first term in Eq. (\ref{aterm}) vanishes,
leaving
\be
A_0 = - \frac{1}{\sqrt{2\sigma}} {\bar D}_{\bar\rho}  \bar{W}(\bar{\rho})
\ee
while for bilinears (B-terms), it is $-m_{3/2}$ yielding the familiar
supergravity relation $B_0 = A_0 - m_0$.

Gaugino masses require in addition a non-trivial dependence of the gauge kinetic function
on the modulus, $\rho$.  This dependence is generic in most of the models of $N = 1$ supergravity derived from extended supergravity and string theory \cite{Ferrara:2011dz}.
The supergravity Lagrangian terms of interest include
\begin{eqnarray}
& -\frac{1}{4} ({\rm Re}\, h_{\alpha \beta})F^{\alpha}_{\mu \nu}F^{\beta \mu\nu}
+\frac{i}{4}({\rm Im}\, h_{\alpha \beta})\epsilon^{\mu\nu\rho\sigma}F^{\alpha}_{\mu \nu}F^{\beta \rho\sigma}
&
 \nonumber \\
&
+\left(\frac{1}{4}e^{G/2} {h_{\alpha\beta}^*}_{\bar{n}}G^{k {\bar n}}G_k\lambda^{\alpha}\lambda^{\beta}+h.c.\right) .
\label{gaugino}
 \end{eqnarray}
For a suitable choice $h_{\alpha \beta} =h(\rho) \delta_{\alpha \beta}$,
one can generate universal gaugino masses
\be
m_{1/2} = \frac{\sqrt{2\sigma}}{6} D_\rho W(\rho) \ln ({\rm Re}\, h^*)_\rho \  .
\ee
In addition to the above expressions, loop contributions may also play 
a role (vital in the case of the KL model).
These expressions have been derived in greater generality, e.g.
when SM Yukawa couplings are also allowed to be moduli dependent,
in \cite{Choi:2004sx,Choi:2005ge}. 

Next, we will compare the resulting soft supersymmetry breaking terms found in the KKLT and 
KL models. We note first that independent of our choice of superpotential,
the soft scalar masses are always $m_0^2 = m_{3/2}^2$. That is,
the chiral multiplets will always be split by the gravitino mass independent
of the magnitude of $D_\rho W$.  

In contrast, both $A_0$ and $m_{1/2}$ are proportional to $D_\rho W(\rho)$
and this differs greatly between the two classes of models:
\begin{eqnarray}
D_\rho W(\rho) & =  & {3\sqrt 2\over a \sqrt \sigma_{0}}\ m_{3/2} \qquad \qquad {\rm KKLT} \nonumber \\
D_\rho W(\rho) &= & 6 \sqrt{2\sigma_{0}}\,  \frac{m_{3/2}}{m_\sigma} \ m_{3/2} \qquad {\rm KL} 
\end{eqnarray}
While the KKLT model could produce a pattern of soft scalar and gaugino masses
all of order $m_{3/2}$ with acceptable phenomenologies \cite{Choi:2004sx,
Choi:2005ge,Endo:2005uy,Choi:2005uz,Falkowski:2005ck,Lebedev:2006qc}, in all models with strong volume modulus stabilization, such as KL models, at the tree-level
one finds that the gaugino masses and $A$-terms are suppressed by $m_{3/2}/{m_\sigma}$
and as such effectively vanish. 

Recall that with $A=B =1,\,a = 0.1,\,b = 0.05$, we have $m_{3/2}/m_\sigma \sim 10^{-13}$. As a result, in the strongly stabilized models, we are driven towards models resembling
those mediated by anomalies \cite{anom}, where the dominant contributions
to gaugino masses and $A$-terms arise from loop corrections and give \cite{Choi:2005ge}
\be
{m_{1/2}} = \frac{b_a g_a^2}{16 \pi^2} \frac{F^C}{C_0}
\label{anomino}
\ee
and 
\be
A_{ijk} = -\frac{\gamma_i + \gamma_j + \gamma_k}{16\pi^2} \frac{F^C}{C_0} \ .
\ee
Here $b_a = 11, 1, -3$ for $a=1,2,3$ are the one-loop beta function coefficients, $\gamma_i$ are the anomalous
dimensions of the matter fields $y_i$ and
\be
 \frac{F^C}{C_0} = - \frac{1}{3} (\ln (\rho + \bar{\rho}) \sqrt{(\rho + \bar{\rho})} D_\rho W)_\rho + m_{3/2}
\ee
is related to the conformal compensator. This is clearly of order $m_{3/2}$.
Note that unlike the case of mixed modulus-anomaly mediation, there is 
no mirage unification in the KL model \cite{Choi:2005uz}.

Because of the loop suppression factor in Eq. (\ref{anomino}),
in order to have a phenomenologically viable model in the context of KL stabilization, 
we are forced to consider relatively large ($\mathcal{O}$(100) TeV) gravitino masses 
in order to have suitably large gaugino masses. Thus we are led to a rather
unique pattern of sparticle masses.  While anomaly mediation plays an important
role in the pattern for gaugino masses, the large soft scalar masses generated from Eq. (\ref{scalar})
yield a spectrum more reminiscent of split supersymmetry \cite{split}.
Indeed, the problem of tachyonic scalars normally associated with anomaly mediated
models is absent here. While this problem is normally alleviated by 
adding a constant mass squared to the anomaly contribution, here we have a direct source
and explanation of the this term. We also note that if the gravitino and scalar masses 
were pushed to very high values ($\gtrsim 10^{10}$ GeV) as in some models of split supersymmetry,
anomaly mediation would be unnecessary.

Gravitinos in the mass range $\mathcal{O}$(100) TeV usually do not pose significant cosmological problems because they decay early. In general, they could be harmful if their decays produce many light supersymmetric particles \cite{Nakayama:2010xf}, but even this problem can be avoided in the simplest  models of chaotic inflation based on the KL construction, where the reheating temperature is small, which strongly suppresses the gravitino production \cite{klor}. 

The heavy sparticle spectrum associated with split supersymmetry
presents a challenge for detection of supersymmetry at the LHC. 
Nevertheless there is an upper limit on $m_0 = m_{3/2}$ of about 100 TeV stemming from the upper bound on 
the Higgs mass from the LHC \cite{higgs}.
An observation of a dark matter in direct
detection experiments with no corresponding signal at the LHC could be an indication pointing
to models such as the one described here.

We also note that the prediction that $B_0 \sim -m_{3/2}$ with the significantly smaller 
smaller $A$-terms can lead to difficulties in finding a consistent electroweak vacuum \cite{drees,vcmssm}.
Furthermore, the large mass scales associated with split supersymmetry also
presents a challenge in finding a consistent electroweak vacuum,
i.e. solutions with $\mu^2 > 0$, where $\mu$ is the Higgs mixing parameter.
However, the former problem may be corrected in a variety of ways, e.g. by adding a mixing term 
to the \K\ potential \cite{gm,cm,Falkowski:2005ck}, while the latter may be
corrected in models containing a right handed neutrino~\cite{cmv,ko}, in models where the input supersymmetry breaking scale is above the GUT scale~\cite{cmv,emo}, or in models with non-universal Higgs
masses \cite{nuhm}.  We will return to these issues elsewhere.

The specific pattern of mass we obtained with the KL potential can also
arise in models where the uplifting is provided by hidden sector F-terms
\cite{Lebedev:2006qc,Fterm}.  In this case, vacua with spontaneously
broken supersymmetry, a small positive cosmological constant and a
hierarchically small  $m_{3/2}$ can
be obtained. This procedure leads to ``matter dominated" supersymmetry
breaking, with the
modulus contribution being suppressed. The resulting soft masses are also
characterized by heavy scalars, though the 
gauginos are lighter than the scalars, they do receive 
contributions from the KKLT uplift in addition to AMSB and are not 
necessarily as light as we have found in the KL model.
The split supersymmetry mass pattern may also appear in models with D-term uplifting, see e.g. \cite{Dudas:2005vv}, and in M-theory on manifolds of G2 holonomy \cite{Acharya:2007rc}.

A distinguishing feature of the models investigated in our paper is that we studied supersymmetry breaking which follows from the simplest versions of the KKLT/KL construction without any additional F-term or D-term contributions. We
particularly emphasized the KL-type models with strongly stabilized internal dimensions, which ensures vacuum stability during inflation.  In general, one can extend such models  to achieve uplifting in the context of $N = 1$ supergravity, see e.g. \cite{Kallosh:2006dv}. 
Then the value of $D_\rho W$ will receive contributions of order the gravitino mass $m_{3/2}$. One may also introduce `split uplifting' models, where part of the uplifting appears due to string theory effects in D = 10, and part is due to N = 1 supergravity. This could allow one to continuously interpolate between models with $D_\rho W = O(m_{3/2})$ and the split supersymmetry models discussed above. However, all such models would be more complicated than the models studied in our paper. Moreover, this would re-introduce the cosmological moduli problem. We were only able to resolve this problem because we did not need to include any light fields in the hidden sector of the models with strongly stabilized moduli.

 \section{Conclusions}
 
In a Minkowski vacuum in N = 1 supergravity, in the absence of a D-term, supersymmetry breaking is determined by $D_\rho W$, or, equivalently, by the gravitino mass $m_{3/2}$. Meanwhile in phenomenological models inspired by string theory, $m_{3/2}$ and $D_\rho W$ are not directly linked to each other, because of the effects of uplifting. 
In the simplest versions of the KKLT model these two parameters are nevertheless of the same order of magnitude. However, as we have shown, the situation is very different in the KL-type models with strong stabilization of the volume modulus, where $D_\rho W \ll m_{3/2}$. These two classes of models have very different phenomenological 
consequences.

In both classes of models,  chiral multiplets are split by the gravitino mass
and $m_0^2 = m_{3/2}^2$.  On the other hand, gaugino masses and $A$-terms
are very different as they are both proportional to $D_\rho W$.  In the simplest KKLT models, 
one can interpolate between modulus mediation and anomaly mediation
to transmit supersymmetry breaking.  The dominant contribution will depend on 
the details of the KKLT superpotential, e.g. on the value of $a \sigma_0$, 
smaller values of  $a \sigma_0$ allowing modulus mediation, while larger
values corresponding to anomaly mediation. In models with a heavy modulus,
$D_\rho W$ is effectively vanishing and one is driven toward anomaly mediation
to generate gaugino masses and $A$-terms. 

However unlike typical anomaly mediation models, here we obtain a massive scalar
sector. Therefore the common problem of tachyonic staus associated
with anomaly mediation is absent here. Instead our spectrum resembles that of split supersymmetry 
with scalar masses approaching the PeV scale.  
Of course there are other immediate consequences of heavy moduli:
There is no cosmological gravitino problem.  Because we are forced to anomaly mediation
to generate gaugino masses, the gravitino must be relatively heavy (approaching the PeV scale for
scalar masses). These gravitinos decay harmlessly, well before nucleosynthesis. Moreover, their production can be strongly suppressed in the simplest models of chaotic inflation based on the KL scenario.
There is no moduli or Polonyi problem because there is no need for a Polonyi field
as supersymmetry breaking is encoded in the uplifting potential. The field $\rho$ does
not create any other cosmological difficulties. Indeed having a successful
model of inflation in supergravity without requiring $H < m_{3/2}$ and  free from the destabilizing effects led us toward heavy moduli in the first place.  

To summarize, cosmological considerations, including the requirement of vacuum stability during inflation and the possibility of solving the cosmological moduli and gravitino problems in string theory and supergravity, point towards models with strongly stabilized internal dimensions, resulting in split supersymmetry with anomaly mediation.

The authors are grateful to S. Dimopoulos, M. Dine, E. Dudas, P. Graham, S. Kachru, R. Kallosh, and T. Rube for enlightening discussions.  The work by A.L. was supported by NSF grant PHY-0756174.
The work of Y.M. Is supported by the French
ANR TAPDMS {ANR-09-JCJC-0146} and the Spanish MICINNs consolider
ingenio 2010 program Under grant `Multi-dark' {CSD2009-00064}.
The work of K.A.O. was supported in part by DOE grant DE-FG02-94ER-40823 at the University of Minnesota.

\section*{Appendix: Generalized KL models with heavy moduli}

Here we will give details of the derivation of Eq. (\ref{KLDW})
which are completely independent of the specific form of the superpotential.

Let return to the scalar potential for the modulus neglecting SM fields.
In terms of the superpotential $W(\rho)$, we can write
\be
V = \frac{1}{6\sigma^2} W_\rho (\sigma W_\rho - 3 W)
\ee
for the potential along the ${\rm Re}\, \rho$ direction. 
The mass of the modulus can be easily computed
\begin{eqnarray}
m_\sigma^2 & = & \frac{1}{2}G^{\rho\bar{\rho}} \frac{\partial^2 V}{\partial \rho \partial \bar{\rho}} \nonumber \\
& = & \frac{1}{18 \sigma^3} ( \sigma^2 W_{\rho,\rho} - 3W ) (4 \sigma^2 W_{\rho,\rho} - 3W) \ .
\end{eqnarray}
Recall also that at the minimum,
\be 
m_{3/2}^2 =  \frac{1}{8 \sigma^3} W^2 \ .
\ee
The limit $m_{3/2}^2 \ll m_\sigma^2$ amounts to
\be
W^2 \ll \frac{4}{9}  ( \sigma^2 W_{\rho,\rho} - 3W ) (4 \sigma^2 W_{\rho,\rho} - 3W ) \ ,
\ee
which is equivalent to $W \ll \sigma^2 W_{\rho,\rho}$.
Note that this was precisely the limit used to approximate Eq. (\ref{KLDW}),
and in this limit
\be
m_\sigma^2 = \frac{2}{9} \sigma  W_{\rho,\rho}^2 .
\label{wrr}
\ee
Finally, we can evaluate $D_\rho W $ from Eq. (\ref{KLDW}) inserting $\Delta \sigma$
\be
D_\rho W = \frac{6 V_{\rm AdS}}{W_{\rho,\rho}} = \frac{18m_{3/2}^2}{W_{\rho,\rho}}\ .
\ee
Substituting $W_{\rho,\rho}$ from Eq. (\ref{wrr})
we arrive at
\be
D_\rho W = 6 \sqrt{2\sigma_{0}} m_{3/2} \frac{m_{3/2}}{m_\sigma} ,
\ee
where $\sigma$ is to be evaluated at the pre-uplifted minimum ($\sigma_0 + \delta \sigma$)
in the notation of the previous section.
Thus, we have shown that in general, 
\be 
D_\rho W  \ll m_{3/2}
\ee
whenever the volume modulus is strongly stabilized, so that $m_\sigma \gg \sqrt{\sigma_{0}} m_{3/2}$. Our result is not specific to the KL model.

\end{document}